# Slowing polar molecules using a wire Stark decelerator

Adela Marian*, Henrik Haak, Peter Geng, and Gerard Meijer

*Fritz-Haber-Institut der Max-Planck-Gesellschaft, Faradayweg 4-6, D-14195 Berlin, Germany*

**We have designed and implemented a new Stark decelerator based on wire electrodes, which is suitable for ultrahigh vacuum applications. The 100 deceleration stages are fashioned out of 0.6 mm diameter tantalum and the array's total length is 110 mm, approximately 10 times smaller than a conventional Stark decelerator with the same number of electrode pairs. Using the wire decelerator, we have removed more than 90% of the kinetic energy from metastable CO molecules in a beam.**

Over the last decade, neutral polar molecules were successfully slowed down by Stark deceleration and subsequently loaded into electrostatic/dynamic traps or storage rings [1]. The technique of Stark deceleration relies on a loss of kinetic energy which is brought about by a coerced gain of Stark potential energy in an inhomogeneous electric field. The output of a Stark decelerator is a packet of state-selected molecules with a narrow velocity distribution and a well-defined tunable velocity. The resulting slow molecules can then be used in a multitude of experiments ranging from spectroscopy and metrology to quantum information and cold collision studies [2-5]. Here, we introduce a new Stark decelerator, where the electrodes are made out of tantalum wires, and we demonstrate its operation by decelerating a beam of metastable CO molecules. The significantly lower costs associated with manufacturing and operating the wire decelerator as compared to a conventional one, the high electric fields achieved within it, and its compatibility with ultrahigh vacuum (UHV) make it attractive for a variety of future applications, in particular trapping experiments and cold-collision experiments.

A schematic overview of the vacuum system is presented in Fig. 1. The molecular beam machine is composed of three differentially pumped vacuum chambers. The first two of them, the source chamber and the small intermediate chamber, are machined out of one 204 mm cubic block of stainless steel and are separated by a thin wall. In this new design, the cube dimensions are ultimately limited by the size of the pump needed for the source chamber, leading to a very compact beam machine. Additionally, the absence of welded joints ensures a good end vacuum. In our setup, the two skimmers delimiting the intermediate chamber have diameters of 2 and 0.5 mm, respectively. The first skimmer is located about 20 mm downstream from the valve nozzle, the second is situated 71 mm behind the first, and both can be independently translated perpendicular to the molecular beam axis.

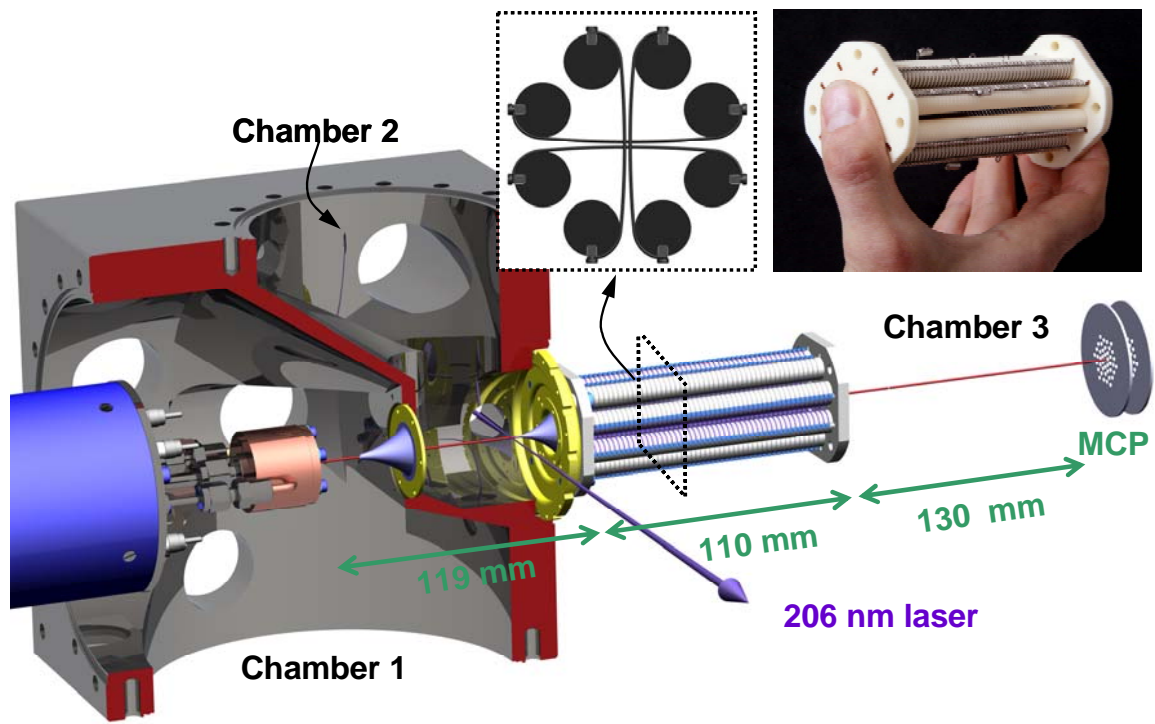

**Figure 1**: Schematic of the experimental setup, consisting of three differentially pumped vacuum chambers. In front of the second skimmer, the ground-state $^{13}$CO molecules leaving the pulsed valve are state-selectively excited to a metastable level using 206 nm light. After passing through the 110 mm long wire decelerator, the entire arrival time distribution of metastable CO is directly detected with a microchannel plate detector. The left boxed inset is a view of the 0.6 mm tantalum electrodes at the exit of the decelerator, looking along the molecular beam axis. The inset on the right is a photograph of the hand-held decelerator.

The third vacuum chamber houses the wire decelerator, which is located 28 mm behind the second skimmer, and consists of an array of 100 equidistant electric field stages. For each stage, high voltage is applied to two parallel 0.6 mm diameter highly polished tantalum wires centered 1 mm apart, leaving a 0.4 mm opening for the molecular beam. As seen in the boxed inset, the 200 wires are bent across 8 ceramic rods and carefully positioned in precut grooves. Successive pairs of electrodes are rotated relative to each other by 90° and are placed at center-to-center distances of 1.1 mm. This adds up to a total length of 110 mm, approximately 10 times smaller than a conventional Stark decelerator with the same number of electric field stages. The positioning of the wires is accurate to about 0.01 mm and was verified by non-contact optical measurements. The 0.4 x 0.4 mm$^2$ opening for the molecular beam makes it possible to employ relatively small voltages (±2 kV) compared with the large decelerators (±20 kV), in order to achieve the same electric field strengths. Generating higher fields is therefore not as technically challenging as for larger-scale machines. We are able to apply up to ±4 kV to the wire electrodes, resulting in maximum electric fields of 200 kV/cm on axis. This decelerator could thus be used for studies of molecules that do not have a large dipole moment or that have a quadratic Stark shift. Additionally, in contrast with existing decelerators, the wire decelerator is designed to be bakeable and can therefore be employed in ultrahigh vacuum. The base pressure in the decelerator chamber becomes 3.5 x 10$^{-10}$ mbar after a bakeout.

The wire decelerator was tested using $^{13}$CO molecules (chosen for experimental convenience) in the low-field-seeking levels of the metastable $a^3\Pi_1$ (v'=0, J'=1) state. First, a pulsed beam (10 Hz) of ground-state ($X^1\Sigma^+$, v''=0) CO molecules is produced in the experiment, by

4expanding a 2 bar mixture of 20% CO in Xe into vacuum through a solenoid valve held at room temperature. Upon exiting the valve nozzle, the most probable velocity of the molecules is about 380 m/s. Then, just before the second skimmer, the molecules are pumped into the metastable state by direct laser excitation at 206 nm using a pulsed dye laser (5 ns, 150 MHz bandwidth, 0.8 mJ/pulse). About 30 mm downstream from the laser excitation point, the molecular packet reaches the first stage of the wire decelerator and, after passing through all the stages, the molecules impinge on a standard microchannel plate (MCP) detector with an active area diameter of 18 mm. Metastable CO molecules can be directly detected thanks to the 6 eV of internal energy they are endowed with, which allows for the entire time-of-flight distribution of the molecules to be recorded as a single trace after each experimental cycle (i.e., every 100 ms). Apart from the high densities obtained with CO in a molecular beam, this direct detection makes metastable CO an attractive choice for proof-of-principle experiments in which accurate measurements of molecular beam velocities are desired [6-8]. As the distance from laser preparation to the exit of the decelerator is 140 mm, which is very short for a molecular beam machine, this compact vacuum system is ideally suited for experiments with short-lived molecules. For metastable CO the radiative lifetime of the $a^3\Pi_1$ (v'=0, J'=1) level is only 2.63 ms [9], in which case a short decelerator is a real advantage.

In a Stark decelerator, polar molecules are slowed down by switching between two static configurations of the electric field. The position of a molecule at the moment when the fields are switched is denoted by a phase angle, and is proportional to the kinetic energy lost per deceleration stage. The same principle of operation is also used for the wire decelerator, however in this case the fields need to be switched 10 times faster due to its reduced dimensions. This is easily realizable in the lab by employing fast semiconductor switches, which have rise times around 50 ns. For a typical measurement at a phase angle of 50°, the fields are switched every 3 μs at the beginning of the deceleration process and then gradually more slowly, every 5 μs or so, once the molecules have lost most of their initial kinetic energy. The entire switching sequence takes around 350 μs, depending on the voltages used. The higher the voltages applied for a given phase angle, the larger the amount of kinetic energy removed per deceleration stage, and the longer the switching sequence. As a direct consequence of the fast switching, the decelerations experienced by molecules in the wire decelerator are on the order of 500 km/s$^2$. Note that the 6D phase-space acceptance of the decelerator operated at a phase angle of 0° for CO molecules is only 300 mm$^3$ (m/s)$^3$, consistent with its small size and approximately 300 times smaller than in a conventional Stark decelerator.

For the measurements shown in Fig. 2, the computer-controlled timing sequence is chosen to slow down metastable $^{13}$CO molecules that enter the wire decelerator at a velocity of 370 m/s, by applying ±4 kV to the tantalum electrodes. Figure 2(a) presents the arrival time distribution of the molecules after they have been gently focused through the decelerator by simultaneously charging all the wire electrodes to ±1 kV. This measurement calibrates the initial beam velocity distribution. The molecules hit the MCP detector after approximately 0.7 ms, corresponding to the 270 mm distance travelled from the laser excitation point. Figure 2(b) gives the complete time-of-flight distribution, which includes the undecelerated part of the beam and the decelerated peak arriving after 0.86 ms for a phase angle of 30°. Figure 2(c) shows a zoom-in of the decelerated peak at three different final velocities. The phase angles used for these measurements are 30°, 50°, and 70°, resulting in a kinetic energy loss per deceleration stage of 0.7 cm$^{-1}$, 1.1 cm$^{-1}$, and 1.4 cm$^{-1}$ respectively. As expected, the arrival of the molecules is progressively shifted to later times. At the same time, using a higher phase angle reduces the number of molecules that can be decelerated. In principle, the final velocity can be tuned to any value, including standstill, which would however prevent the molecules from reaching the detector. In the current experimental arrangement, a clear peak is still observable for a final velocity of 99 m/s (not shown), corresponding to 1.84 ms time of flight. In this case, more than 92% of the kinetic energy of the metastable CO molecules has been removed. However, it is increasingly harder to observe molecular packets decelerated to even lower velocities for two reasons: (i) transverse expansion of the slow beam, and (ii) arrival times approaching the radiative lifetime of the molecular state.

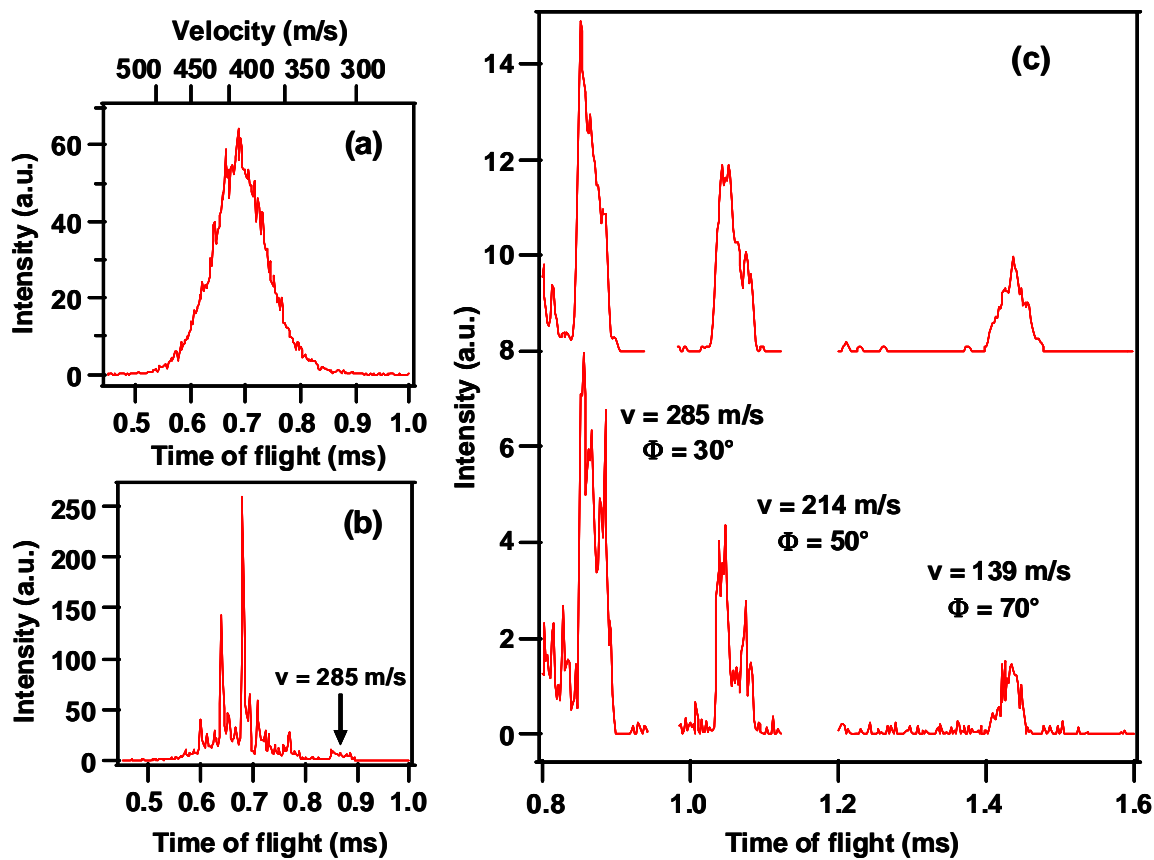

**Figure 2**: Arrival time distributions of the metastable $^{13}$CO molecules on the MCP detector. (a) Measured initial distribution of the molecules. The velocity of the molecules is shown on the top axis, while the time scale on the bottom axis is given with respect to laser preparation. (b) Time-of-flight profile for a phase angle of 30°, resulting in deceleration from 370 m/s to 285 m/s. (c) Zoom-in on three deceleration peaks recorded for increasingly higher phase angles and lower final velocities, as indicated on the graph. The upper curves are results of three-dimensional trajectory simulations for the parameter sets used in the experiment. All experimental traces are shown on the same vertical scale and have been averaged for 1 hour at 10 Hz.

Both these problems can be easily remedied by minimizing the distance between the exit of the decelerator and the MCP detector.

The experimental time-of-flight profiles are in good agreement with the corresponding simulated profiles resulting from three-dimensional trajectory calculations, which are shown with a vertical offset above the measured data in Fig. 2(c). As a remark, the undecelerated part of the beam is not so well reproduced in the simulations, which could be attributed to nonadiabatic transitions [10]. It is important to note that the pressure in the decelerator chamber remains in the $10^{-10}$ mbar range during the experiment, at a value of 7 x $10^{-10}$ mbar, ensuring operation in ultrahigh vacuum.

After this first use of tantalum wires for Stark deceleration, further improvements in wire design, detection efficiency, signal intensity, and end vacuum are immediately foreseeable. Moreover, a mode-matched electrostatic or AC electric trap [1] can be integrated in the future at the exit of the decelerator. One very attractive feature of the wire decelerator is its suitability for UHV experiments, making it a promising source of slow molecules for a variety of atomic and molecular physics applications. In particular, the wire decelerator was developed for an experiment that aims to produce ultracold molecules via sympathetic cooling with ultracold atoms using spatially overlapped traps. The idea and experimental setup have been described elsewhere [11]. The compact ultrahigh-vacuum molecular-beam machine presented here can accommodate all the spatial and vacuum requirements imposed by the existing atom apparatus. Thus, apart from the already mentioned possibility to trap the packets of slow molecules produced by such a wire decelerator, the door is also open for collision experiments between slow molecules and trapped cold atoms.

We are indebted to S. Y. T. van de Meerakker for use of his simulation program and valuable suggestions. We gratefully acknowledge fruitful discussions with S. A. Meek, L. Scharfenberg and J. J. Gilijamse. This research was supported by a Marie Curie Intra-European Fellowship within the 6th European Community Framework Programme.

* Electronic address: marian@fhi-berlin.mpg.de